# Protection of genomic information: a classical and a quantum approach.

Guy Dodin

Université Paris Diderot, Laboratoire ITODYS-CNRS UMR 7086, 15 rue Jean-Antoine de Baïf 75005 Paris
dodin@paris7.jussieu.fr
dodin@biologie.ens.fr

Abstract

Splitting a literal genomic sequence into 4 binary files is enough to ensure *confidentiality* and *integrity* during storage and transfer of information. The binary files are resources for RSA or one-time-pad (OTP) cryptography protocols.

It is speculated that representing nucleic bases as Bell's states in a 'quantum' view of a sequence would provide tools for genomic data protection when implemented in an authentic quantum computer, soon to come as a practical and readily available device.

Secure storage and distribution of data from sequencing genomes of bacteria, viruses and humans (with special emphasis on the latter for its significance in personal medicine) is a major issue.

Here it is shown how the genomic sequences can be safely encrypted and transferred and how tools developed for this purpose can be extended to the protection of any type of data. A venture into a 'quantum' view of a genomic sequence may provide formal means for safe genomic data protection in the future.

**Security and integrity of genomic sequence storage**

The level expected for security of the genomic information (or for any other data or items) results from a permanent trade-off between the value attributed to the object to be protected and the efforts and resources an attacker would engage to break confidentiality. Hence the subsequent protocols for data storage and transfer are occasional, not meant to provide absolute protection but are likely to make it difficult for malicious actions to be successful.

*1-a classic sequence*

Nucleic acid sequences, as represented by strings of four letters, can be viewed as vector-like objects in a four-dimension space endowed with four *orthogonal* basis vectors. The projection of the sequence onto one dimension can be expressed as a binary string. For example, projecting sequence –---AGTCAAG--- leads to a 4-row array, with yhe rows referred to as Bin(base).

A  –-1000110---   Bin(A)
T  –-0010000---   Bin(T)
G  –-0100001---   Bin(G)
C  –-0001000---   Bin(C)

When *separately* stored, the Bin(base) files offer safe protection of genomic data. Indeed, fraudulent access to one file makes it hardly possible to unveil the full nucleotide sequence since the number of combinations formed with the remaining three letters is $3^N$ for a *N*-base long DNA string. Even if two Bin(base) files have been hacked, the probability of recovering the genuine sequence remains as low as $(2^N)^{-1}$. Access to three files ($1^N$ possible remaining combination) is enough to retrieve the genuine nucleotide sequence, provided *none of the binaries is corrupted*.

In addition to *confidentiality*, the binary splitting allows to detect breaches in i*ntegrity* if an attacker modifies the length or the compositions of the Bin(base). Indeed, all files should have same length and, *only one digit '1'* should appear in each column in the 4xN array representing the sequence. The splitting into binaries, the reconstruction of the literal sequences as well as the control of integrity is readily achieved using a simple algorithm (a Python code has been develop to this respect).[1]

Straightforward conversion of the binary *strings into numbers* (binary, decimal, hexadecimal) readily provides numerical resources of interest for general purpose cryptography. For example, the third binary file necessary for reconstructing the genuine genomic sequence can be encrypted using the RSA algorithm with primes closest to numbers from the two other binary strings.

Another safe encryption tool is the one-time-pad algorithm (OTP). However, OTP is *provably secure* only if the message to be encrypted is Xored (exclusive OR logical operation) *only once* with an *authentic random binary sequence*. True random sequences cannot be provided by classical algorithms, deterministic in nature but the rapid development of quantum computers whose logic is based on the physical principles of quantum mechanics, is likely to provide genuine randomness. Nevertheless, an acceptable level of security can be reached using binary pads derived from a random access to the immense resources provided by genomic data banks ([ftp://ftp.ncbi.nlm.nih.gov/genomes/refseq/bacteria/;](ftp://ftp.ncbi.nlm.nih.gov/genomes/refseq/bacteria/) [ftp //ftp.ensembl.org/pub/release](ftp://ftp.ensembl.org/pub/release)99/fasta/homo_sapiens/dna/)

Patterns and regularities in genomic sequences, barring them as good candidates for OTP encryption, are tentatively detected from the Fourier transform (FT) of the binary files. In this respect, compact, intronless bacterial or viral genomes should be excluded as they present a remarkable peak at frequency 1/3 in the Fourier spectrum, resulting from the repetition of codons. This pattern is not observed in the human genome (where only 2-3 % correspond to gene coding) nor are significant regularities in the low frequency region of the spectrum. (Figure 1)

In the course of the import from the data bank, an attacker might gain knowledge of the genomic sequence and the chromosome it originates by 'blasting' the sequence against the whole data bank. Hence, splitting the sequence into its Bin(base) prior to transfer would insure enhanced security. Alternatively, safe transfer of a *literal* sequence might be achieved in a close future using quantum devices. Several platforms are readily accessible for *simulating* a quantum computer on a classical machine thus permitting the development of quantum algorithms in the meantime.[2] (IBM's [Qiskit](Qiskit), Microsoft's Q#, Qutyp quantum toolbox in Python).

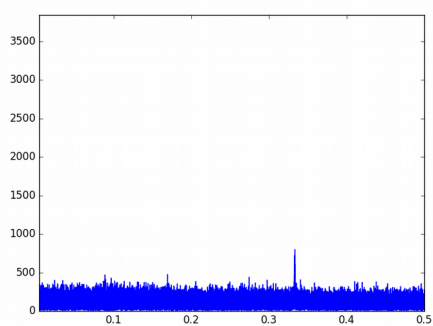
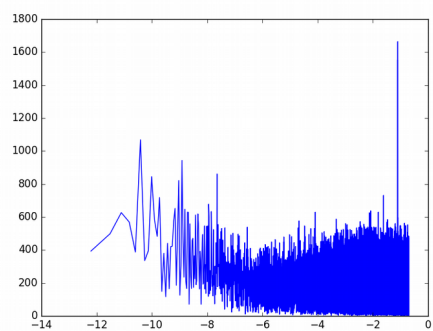
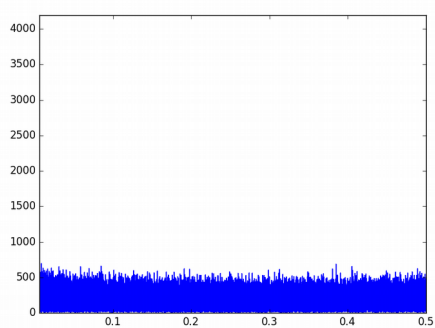
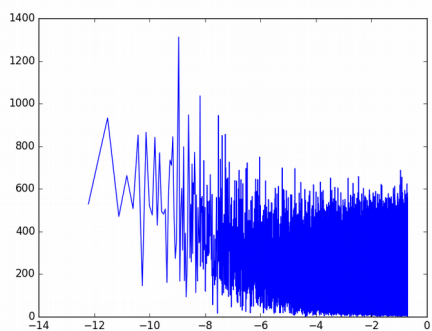

**Figure 1:** from upper left: Fourier Transform (FT) of *Bin(A)* of a random 200kB-long bacterial sequence (here E.coli, K13_MG1655); Log plot; Fourier Transform of Bin(A) of a random 200kB-long human genome sequence (here Human chromosome 21,GRCh38); Log plot of c.

2-a *'quantum' sequence*
In a 'quantum' genomic sequence, *a given position* is no longer occupied by a well-defined nucleotide but rather consists in a superposition of the four building bases. The 'classical' sequence would only materialized when *observed* and, in the process (the measurement), all contributions in the superposition would collapse, except that observed. Although real genomic sequences are not quantum objects (nor is Schronfiger's cat!), the 'quantum' approach would tentatively lead to a fresh view on a DNA sequence.

The Dirac formalism, the general tool to describe quantum states is adopted here. Conventionally, a state S is represented by a 'ket' vector, |S> associated with a column matrix. The transpose of the 'ket' matrix is a row matrix associated with the 'bra' vector <S|.[3-4]

Accordingly, the state at a given position on the sequence can be represented by the ket:

$|\psi> = \alpha|A> + \beta|T> + \gamma|G> + \delta|C>$ (1) or,

$|\psi> = \alpha|0> + \beta|1> + \gamma|2> + \delta|3>$ where numbers stand respectively for bases. The coefficients (Greek letters) are complex numbers. $|\psi>$ will be referred to as a *ququart* (4 contributions in the representation)[5].

The scalar product of $|\psi>$ by itself (expressed as $<\psi|\psi>$) represents the probability of a position on the sequence to be occupied by any of the 4 bases and should be equal to 1, consistently with the base vectors being orthogonal.

$\alpha^2 + \beta^2 + \gamma^2 + \delta^2 = 1$

$|\psi>$ can be represented as a column matrix whose elements are the coefficients in the base $|0> .... |3>$. Given equal probability for each base to occupy one position, the coefficients are likely to be equal to 1/2. The probabilistic nature of the coefficients will be further underlined and discussed.

In the course of this contribution column matrices will be represented as the transposes T of row matrices for sake of easier writing.

$|\psi> = 1/2[1\ 1\ 1\ 1]^T$
$|0> = [1\ 0\ 0\ 0]^T$
$|1> = [0\ 1\ 0\ 0]^T$

$|2\rangle = [0\ 0\ 1\ 0]^T$
$|3\rangle = [0\ 0\ 0\ 1]^T$

Although straightforward, the ququart representation suffers significant drawbacks. First, the information in $|\psi\rangle$ can only be retrieved upon *measurement* which would lead to *statistical*, information-poor results. For instance the probability that $|\psi\rangle$ is in state $|0\rangle$ is measured by the square of the scalar product $\langle 0|\psi\rangle$ and is 1/4 as could have been anticipated. Would it be possible to create, starting from $|\psi\rangle$, states which, when measured, would deliver a *unique* result? This might be achieved by applying to $|\psi\rangle$ various operations (represented by the actions of 4x4 square matrix operators) at the price formal complexity. Second, and most concerning, developing ququart-based algorithms and protocols may be pointless given the formidable challenge of manufactoring stable, reliable and readily available concrete *physical* devices.

Alternatively, a ququart state can be represented as a combination of 2 *qubits*. Qubits are formally easy to manipulate and an abundant literature is devoted to their use in quantum protocols. Moreover, qubits have already materialized as concrete physical set-ups implemented into true quantum computers.

*Qubit*

A qubit is represented as the superposition of 2 elementary states:
$|\psi\rangle = \alpha|0\rangle + \beta|1\rangle$
$\alpha$ and $\beta$ are complex numbers with $\alpha^2 + \beta^2 = 1$

In contrast to a classic view, $\alpha$ and $\beta$ can take any values conventionally represented as points on the Bloch sphere (figure 2). This quantum indetermination makes qubits eligible as resources for true random numbers generators (see above discussion on OTP).

When coefficients $\alpha$ and $\beta$ are expressed in terms of angles on the Bloch sphere (figure2, a qubit is written as:
$|\psi\rangle = \cos(\theta/2)|0\rangle + e^{i\varphi}\cdot\sin(\theta/2)|1\rangle$

Although undetermined, the most probable qubit would correspond to $|0\rangle$ and $|1\rangle$ equally contributing in the superposition ($\theta = 90°$).
$|\psi\rangle = 1/\sqrt{2}\ (|0\rangle + |1\rangle)$ if $\varphi = 0$

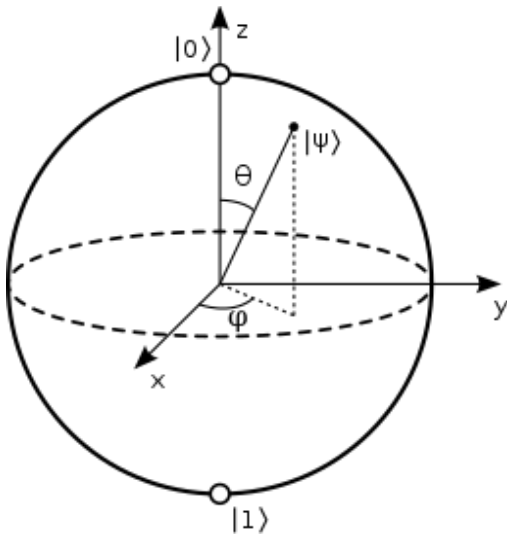

**Figure 2**: representation of a qubit on Bloch sphere.

*Exchanging one ququart for 2 qubits*
In a quantum world, the space of a 2-qubit state is the *tensor product* of the individual spaces. Consequently, the ket |ψ> associated to qubits |ψ$_1$> and |ψ$_2$> would be expressed as the tensor product |ψ$_1$>⊗|ψ$_2$>.
The space of |ψ> has 4 dimensions and hence spans the same-dimension space as a ququart.
|ψ>=|ψ$_1$>⊗|ψ$_2$> = α$_1$α$_2$|0$_1$>⊗|0$_2$>+α$_1$ β$_2$|0$_1$>⊗|1$_2$>+α$_2$ β$_1$|1$_1$>⊗|0$_2$> + β$_1$ β2|1$_1$>⊗|1$_2$>, or, omitting subscripts on basis vectors:
|ψ> = α$_1$α$_2$|00> + α$_1$ β$_2$|01> + α$_2$ β$_1$|01> +β$_1$ β$_2$|11>
|ψ> =1/2 (|00> + |01> + |01> +|11>) with all α and β = 1/√2.
|ψ> is represented as a column matrix:
|ψ> = 1/2[1 1 1 1]$^T$
The basis vectors of the space product are:
|00>=[1 0 0 0]$^T$
|01>=[0 1 0 0]$^T$
|10>=[0 0 1 0]$^T$
|11>=[0 0 0 1]$^T$
Since *two classical bits* (each with value 0 or 1) are necessary and sufficient to encode the nucleic bases (00 would be A, 01 T, …), |ψ> can be formally rewritten as:
|ψ> =1/2 (|A> + |T> + |G> + |C>)
Was the journey into the qubit representation useless as it ends up with the same |ψ> as ququart state (eq.1) suffering the same limitation arising from the statistical outcome of the measurement? Not quite. First, as will be seen below, it is now easy to construct states able to yield, when measured, a *single* result. Moreover, from a practical viewpoint, manufacturing

concrete physical qubits, although still demanding, is in rapid progress and true qubit-based quantum computers are already operating.
From the 4 |ij> basis vectors, 4 special vectors, the so-called Bell's states, *not reducible to tensor products*, can be readily built.[3]

*The Bell's states*
Quantum gates (matrix operators) used to construct the entangled Bell's states correspond to rotations around the axes in the Bloch sphere.
Quantum gates are *unitary operators whose actions are reversible.*[4,7]
The Hadamard gate (H), a single qubit operator, when applied to the basis vectors |0> and |1> of a qubit leads to:
H|0> = $1/\sqrt{2}$ (|0> + |1>) = |ψ$_+$>
H|1> = $1/\sqrt{2}$ (|0> - |1>)| = |ψ$_-$>
The CNOT (cX) gate acting on a *2-qubit system* flips the state of the second qubit only when the first qubit is in state |1>.
Hence, applying first H and then cX onto the basis vectors of |ψ> (|00>, |01>, |10>, |11>) leads to the 4 Bell's states.
For example:
H action on first qubit of |00>:
H|00> = $1/\sqrt{2}$ (|0>+|1>) |0> = $1/\sqrt{2}$ (|00>+|10> then cX acting on second qubit:
cX(H|00>)= $1/\sqrt{2}$( |00> + |11|)= Bell$_{00}$, represented by the column matrix :
[$1/\sqrt{2}$  0  0  $1/\sqrt{2}$]$^T$
With 0-valued elements in its representative matrix, Bell$_{00}$ is not a tensor product (where all matrix elements should be non-zero -see above-). The two qubits are entangled in state Bell$_{00}$.
H then cX applied to the other basis vectors lead to Bell's states Bell$_{01}$, Bell$_{10}$, Bell$_{11}$.
Alternatively Bell's states can be derived from Bell$_{00}$ applying Z, X gates.[7]
The 4 Bell's states are:
Bell$_{00}$ =$1/\sqrt{2}$( |00> + |11>) from H.cX acting on |00>
Bell$_{01}$ =$1/\sqrt{2}$( |01> + |10>)                    |01>
Bell$_{10}$ =$1/\sqrt{2}$( |00> - |11>)                    |10>
Bell$_{11}$ =$1/\sqrt{2}$( |01> - |10>)                    |11>
Now if A wants to inform B that adenine (encoded as 00) is the base at a given position in the sequence, A will consider state |00>, transform it into Bell$_{00}$ and send Bell$_{00}$ to B. B will apply cX to the qubits and *then* H to the first bit (the reverse order was used to construct Bell$_{00}$) and will retrieve state 00, corresponding to adenine, with probability equal to 1. Hence, B will *unambiguously* know that adenine is the base involved. Alternativey, transmission of the information can also be performed through the superdense coding procedure[3] that, however, necessitates the intervention of a *third party* thus introducing complexity, unnecessary in the context of this presentation.

to th A full genomic sequence can progressively be transferred and stored. For example,

$cX(Bell_{00}) = 1/\sqrt{2}( |00\rangle + |10\rangle)$

Then H on first qubit :

$1/\sqrt{2}(1/\sqrt{2}(|0\rangle + |1\rangle)|0\rangle + 1/\sqrt{2}(|0\rangle - |1\rangle)|0\rangle) =$

$1/2 (|00\rangle + |10\rangle + |00\rangle - |10\rangle) = |00\rangle$

Why could not A simply send the message 'adenine ' or A or 00 to B instead of going through the tedious procedure of associating a Bell's state to the base? A classical, non quantum message (like 'adenine', A or 00), can be intercepted by an attacker without B knowing it. If now the eavesdropper intercepts the *quantum* message, he can decode it, as B would, but measurement will collapse the state and B will know the message has been hacked. He can also try to *make a copy* of the message for further processing but the no-cloning theorem[3, 6] prevents any information to be gained from copying a quantum message and the efforts of the attacker remain useless.

In conclusion, the quantum approach appears appealing for secure data storage and exchange. However it is, so far, only a formal view that would become an efficient tool for handling genomic data only if implemented on a true quantum computer. Even if 'quantumized' to some extent by introducing uncertainty into the coefficients, description of states such as ψ> or Bell's would remain deterministic since randomness would originate from pseudo random number generators.

**References**

1-*Dodin, G;* biorxiv, doi:https://doi.org/10.1101/230516
2-https://www.microsoft.com/en-us/quantum/development-kit
https://developer.ibm.com/open/projehcts/qiskit/
http://qutip.org/

3-Nielsen, M, A; Chuang, I (2000) *Quantum Computation and Quantum Information. Cambridge: Cambridge University Press.*

4-Wilde, Mark M, (2017) Quantum Information Theory, Cambridge University Press
5-Iao-Min Hu, Yu Guo, Bi-Heng Liu, Yun-Feng Huang, Chuan-Feng Li, Guang-Can Guo ; arXiv:1807.10452v1
6-*Wootters, William; Zurek, Wojciech (1982); Nature. **299** (5886)*

7-Quantum gates :

$$H = \frac{1}{\sqrt{2}} \begin{bmatrix} 1 & 1 \\ 1 & -1 \end{bmatrix}$$

$$\text{CNOT} = cX = \begin{bmatrix} 1 & 0 & 0 & 0 \\ 0 & 1 & 0 & 0 \\ 0 & 0 & 0 & 1 \\ 0 & 0 & 1 & 0 \end{bmatrix}$$

$$X = \begin{bmatrix} 0 & 1 \\ 1 & 0 \end{bmatrix}$$

$$Z = \begin{bmatrix} 1 & 0 \\ 0 & -1 \end{bmatrix}$$